\documentclass[11pt]{article}
\usepackage{fleqn}
\usepackage[dvips]{graphicx}
\begin{document}

\title
{\bf 
Proposal to observe the strong Van der Waals force in the low 
energy neutron-Pb scattering\footnote{NUP-A-2000-7}
}
\vspace{20mm}
\author{Tetsuo Sawada \\
{\small \em Atomic Energy Research Institute, Nihon University, 
Tokyo, Japan 1010062}
\thanks{Associate member of AERI for research. \ \ \  
  e-mail address: t-sawada@fureai.or.jp}}
\date{}

\maketitle
\thispagestyle{empty}
\vspace{70mm}
\begin{flushleft}
{\large\bf Abstract}
\end{flushleft}

  In the neutron-Pb scattering in the MeV. region, existence of the 
long range interaction has been known and people have hopefully 
expected to understand it
as the effect of the electric polarization of the neutron.     
However the precise determination of $\alpha_{n}$ in the 
very low energy region indicates that the polarizability of 
the neutron $\alpha_{n}$ is around two order of 
magnitude smaller compared 
to what is expected in the phenomena in the MeV. region.  
We need another strong long range potential  which decreases 
faster than $r^{-4}$.   On the other hand, analysis of the 
low energy proton-proton data revealed the existence of the 
strong long range potential whose tail was $V(r) \sim -C/r^ 
{\alpha}$ with $\alpha=6.08$.  The neuton-nucleus scattering 
is a good place to see such a force, since the strength $C$ is 
magnified by factor $A$, the mass number.  Using the long 
range parameters obtained in p-p, we predict the characteristic 
angular dependency of the n-Pb amplitude for fixed energy.  
Appearance of a cusp at $\nu=0$, which points upward, in the 
once subtracted P-wave amplitude $a_{1}(\nu)/\nu$ is  
predicted. 

\newpage
\setcounter{page}{1}

\section{Introduction }

 By the neutron transmission experiments of $Pb^{208}$ in the energy 
 range of $0.05keV < T_{lab} < 40keV$, Schmiedmayer et.al.\cite
 {polarizability} successfully 
 determined the electric polarizability of the neutron $\alpha_{n}$. 
 In the determination of $\alpha_{n}$, they fitted the n-Pb$^{208}$ 
 cross  section as 
\begin{equation}
\sigma(\nu)=11.508(5)+0.69(9) \sqrt{\nu}-448(3) \nu+9500(400) \nu^2
\end{equation}  
where $\sigma$ and $\nu$ are measured in the units of barn and 
fm$^{-2}$ respectively.   The $\sqrt{\nu}$ term arises from the 
long range potential of the electric polarization of the neutron. 
From the coefficient of $\sqrt{\nu}$, they obtained the value 
$\alpha_{n}=(1.20 \pm 0.15 \pm 0.20) \times 10^{-3} \; fm^{3}$. 
On the other hand, when all the forces are short range, $\sigma(\nu)$ 
must be a polynomial of $\nu$.
However the smallness of $\alpha_{n}$ obtained from 
this precise experiment reveals an embarrassing situation.   Namely 
the n-Pb data in the MeV. region have the behavior 
characteristic of the long range force\cite{anikin}, 
and if we attribute it to 
the same electric polarization of the neutron, the value of 
 $\alpha_{n}$ becomes 
 around two orders of magnitude  greater compared to what is obtained 
 in the very low energy region.\cite{pokot}
    Therefore we need another long 
 range force which does not alter the coefficient of 
 $\sqrt{\nu}$.   An example of the required long range force is 
 the strong Van der Waals force, because it gives rise to the 
 singular term $\nu^{3/2}$ or $\nu^{2} \log \nu$ in Eq.(1) as we 
 shall see in the next section.

        Since in the hadron physics, interactions are believed 
 for a long time to be short range arising from the exchanges 
 of a pion, a set of pions or heavier particles, the appearance 
 of the strong Van der Waals force may sound strange.  However 
 in 1960's our view of hadrons changed from elementary particles 
 to composite particles.    In most of the composite model of 
 hadron, the fundamental force is strong or super-strong  
 Coulombic force, and it is responsible for the formation of 
 the "neutral" bound states, and which are identified with the 
 hadrons.  Here "neutral" means total "charge" zero, in which 
 "charge" corresponds to the fundamental Coulombic force.
 As in the case of the ordinary (electric) atoms, the quantum 
 fluctuation gives rise to the Van der Waals interaction 
 between the "neutral" composite particles, namely between 
 hadrons.   When the radius of the hadron and the "fine structure 
 constant" $^{*}e^2$ are known, we can estimate the strength 
 $C$ of the Van der Waals potential
  , at least in the order of magnitude.
 It turns out that when the fundamental Coulombic force is 
 super-strong such as in the case of the dyon model of 
 Schwinger\cite{dyon} 
 , in which $^{*}e^2= 137.04/4$, $C$ becomes large 
 enough to compete with the potential of the one-pion exchange.
 Therefore it is desirable to search for the strong Van der 
 Waals interaction whenever sufficiently precise data are 
 available.   The neutron-Pb scattering has an advantage in the 
 search, because for large $r$ the strength $C$ of the strong 
 Van der Waals potential of the nuclear force is magnified by a 
 factor $A$ in the neutron-nucleus potential, where $A$ is the 
 mass number of the nucleus and we shall set $A=208$ in this 
 paper.   
 
       From the analysis of the S-wave phase shift data of the 
 low energy proton-proton scattering, we have determined the 
 parameters of the long range 
 tail of nuclear potential $v(r) = -C/r^{\alpha}+\cdots$ in which   
 $\alpha=6.08$ and $C=0.196$ in the unit of the Compton wave length 
 of the neutral pion.\cite{ppvdw}
 Once the tail of the nuclear potential is known, it is not 
 difficult to compute the asymptotic behavior of the n-Pb 
 potential and to  determine the singular behavior of the n-Pb 
 scattering amplitude.  Since the singular term has the form 
 $(-t)^{\gamma}$, where $\gamma=(\alpha-3)/2$, we can obseve the 
 anomalous behavior of the n-Pb amplitude in two places.
 One is the anomalous behavior $(1-z)^{\gamma}$ of the amplitude 
 at $z=1$ for fixed $\nu$, and the other is the anomolous term 
 $\nu^{\gamma}$ in the partial wave amplitudes $a_{\ell}(\nu)$. 
 The aim of the present paper is to predict the shapes and the 
 magnitudes of these singular behaviors of the n-Pb amplitude by 
 using the parameters of the long range force determined from 
 the p-p scattering.
 
       In section 2, the relations between the singular behavior 
 of the amplitude at $t=0$ and the asymptotic behavior of the 
 potential at large $r$ are described.   In section 3, the anomaly 
 of the angular distributions of the amplitudes are shown, in 
 which the incident energies are fixed at $T_{lab}=$ 0.25, 0.5, 
 0.75 and 1.0 MeV. respectively.   In section 4, the once 
 subtracted P-wave amplitude $a_{1}(\nu)/\nu$ is computed, and it 
 is shown that it has a characteristic cusp $\nu^{\gamma -1}$ at 
 $\nu=0$.   Section 5 is used for comments and discussions.  In 
 Appendix, the long range components of the nuclear potential 
 obtained from the S-wave amplitude of the p-p scattering are 
 summarized along with the brief explanation of the analysis of 
 the p-p data.

\section{ Asymptotic form of the potential and the \\ 
singular behavior of the amplitude }
 
    When we want to confirm the existence of the long 
 range force unambiguously, it is 
desirable to use the difference of the analytical structure of 
the scattering amplitude $A(s,t)$.    This is because the 
 amplitude $A(s,t)$ is 
regular in the neighborhood of $t=0$ if all the forces are 
short range, on the other hand when the long range force is 
acting an extra singularity appears at $t=0$.    It is 
fortunate that $t=0$ is the end point of the physical region 
 $-4 \nu \leq t \leq 0$, and we can in principle settle the  
ploblem whether the 
strong interaction involves the long range force, when the 
sufficiently accurate data are given.

    Since the potential $v(r)$ and the spectral function $A_{t}(s,t)$ 
of the scattering amplitude are connected by 
\begin{equation}
  v(r)=-\frac{1}{\pi m^2} \frac{1}{r}\int_{0}^{\infty} dt A_{t}(s,t) 
 e^{-r \sqrt{t}} \quad ,
\end{equation} 
the parameters of the threshold behavior of $A_{t}(s,t)$ and those 
of the asymptotic behavior of the potential, which are defined by 
\begin{equation}
A_{t}(s,t)=\pi C' t^{\gamma}+ \cdots \qquad and \qquad v(r) \sim 
 - C \frac{1}{r^{\alpha}}+\cdots
\end{equation} 
respectively, relate to each other by 
\begin{equation}
\alpha=2 \gamma +3 \qquad and \qquad C=- \frac{2 C'}{m^2} 
\Gamma(2 \gamma +2) \quad .
\end{equation}  
From the threshold behavior of $A_{t}(s,t)$, the singular behavior 
of the amplitude $A(s,t)$ at $t=0$ is determined : 
\begin{equation}
A(s,t)=- \frac{\pi}{\sin \pi \gamma} C' (-t)^{\gamma} + \mbox{
(polynomial of $t$ ) .} 
 \end{equation}  
 In particular when $\gamma$ is an integer $n$, we must take the 
 limit $\gamma \rightarrow n$ of Eq.(5), and the singular term 
 becomes $(-1)^{n+1} C' (-t)^{n} \log (-t)$.
  For the case of the 
 strong Van der Waals force, the power $\alpha$ of the asymptotic 
 behavior of the potentials are $\alpha=6$ and  $\alpha=7$ for the 
 London type and the Casimir-Polder type respectively, and which 
 correspond to $\gamma=1.5$ and $\gamma=2$ respectively.  
 
        Merits to 
 use the large nucleus as the target of the neutron scattering are 
 twofold.  When we construct the neutron-nucleus potential by making 
 the convolution of the nuclear potential and the form factor of
 the nucleus, the strength of the long range tail of the potential 
 becomes $A$ times large compared to that of the nuclear potential 
 between nucleons, where $A$ is the mass number of the nucleus and 
 in our case $A=208$.    On the other hand, since the one-pion 
 exchange term of the nuclear potential involves the factor $(\vec{
 \sigma}_{n} \cdot \vec{ \sigma}_{j})$ or $(\vec{
 \sigma}_{n} \cdot \hat{r})( \vec{ \sigma}_{j} \cdot \hat{r})$, sum 
 of the contributions from all the constituent nucleons cancels out.
  Therefore we may expect to get very clear view of the singularity 
 of the long range force at $t=0$, because the spectrum of the 
 two-pion exchange starts at $t=4$ and gives rise to an almost 
 constant back ground in the small neighborhood of $t=0$. Throughout 
 of this paper, we shall use the neutral pion mass and its Compton 
wave length as the units of the energy and the length respectively. 
However because of the large radius $r_{1}$ of the nucleus, eg. for 
$Pb^{208}$ $r_{1}$ is $4.42$, we cannot neglect the back ground 
ploynomial function of $t$, when the domain of $t$ becomes wide and 
lies out side of $\sqrt{-t} < 1/r_{1}$.   
In order to know the necessary degree of the back ground polynomials, 
in section 3 and 4 we shall also compute the n-Pb amplitude 
arising from 
short range potential, the potential of the $\sigma$-meson exchange 
in particular.
 
       Since $t=-2 \nu (1-z)$, the singular term $A^{sing}(s,t)$  
becomes  
\begin{equation} 
 A^{sing}(s,t)=- \frac{\pi}{\sin \pi \gamma} C' 
 (2 \nu)^{\gamma}(1-z)^{\gamma} \quad ,
 \end{equation} 
 and for fixed $\nu$, we must observe the singular behavior 
 $(1-z)^{\gamma}$ in the neighborhood of $z=1$.    Another 
 interesting property of the long range force is the anomaly of the 
 threshold behavior of the partial wave amplitude $a_{\ell}(\nu)$. 
 For the short range force, it is well-known the amplitude $a_{\ell}
 (\nu)$ is proportional to $\nu^{\ell}$ at the threshold.  This is 
 because in the partial wave projection of the polynomial function of 
 $t$, $z^{\ell}$ appears first in the term of $t^{\ell}$, thus the 
 factor $\nu^{\gamma}$ always appears in $a_{\ell}(\nu)$.    On the 
 other hand, the result of the partial wave projection of the 
 singular term is
 \begin{equation}
 a_{\ell}^{sing}(\nu)=\frac{1}{2} \int_{-1}^{1} dz P_{\ell}(z) 
  A^{sing}(s,t)=- \frac{\pi}{\sin \pi \gamma} C' 
  (2 \nu)^{\gamma} I_{\ell}(\gamma) \quad ,
 \end{equation} 
 where the patial wave projection of $(1-z)^{\gamma}$ is
 \begin{equation}
  I_{\ell}(\gamma)=2^{\gamma} \frac{(-\gamma)(1-\gamma) \cdots 
 (\ell -1 -\gamma)}{(1+\gamma)(2+\gamma) \cdots  (\ell +1 +\gamma)}  
 \end{equation} 
 for $\ell>0$ and for $\ell=0$, $I_{0}(\gamma)=2^{\gamma}/(1+\gamma)$.
 Therefore all the partial wave amplitudes have a term whose 
 threshold behavior is proportional to $\nu^{\gamma}$, if the 
 long range force is acting.   For example for the case of the Van 
 der Waals force of the London type ($\gamma=1.5$), all the partial 
 waves other than S and P waves have the threshold behavior 
 $\nu^{1.5}$.  
  Moreover although the threshold behavior of the P-wave is normal 
 and we can consider $a_{1}(\nu)/\nu$, it involves a term proportional 
 to $\sqrt{\nu}$ arising from the singular term  $ A^{sing}(s,t)$. 
 Since Eqs.(7) and (8) indicate that the term of $\sqrt{\nu}$ has the 
 negative sign, when the asymptotic force is attractive ($C'>0$), the 
 cusp of $a_{1}(\nu)/\nu$ at $\nu=0$ must points upward.  The shape of 
 the P-wave in the low energy region is one of the ideal place to 
 observe the effect of the long range force, because for the higher 
 partial waves, we cannot obtain precise data in the low energy region.
  On the other hand, although the cusp of $\sqrt{\nu}$ with opposite sign 
 appears in the once subtracted S-wave amplitude $(a_{0}(\nu)-a_{0}(0))
 /\nu$, extremely precise value of the scattering length $-a_{0}(0)$ 
 is necessary to observe such a cusp.    Therefore in the search of the 
 long range force, we shall concentrate on the singular behavior of the 
 angular distribution in the forward region $z=1$ and on the shape of 
 $a_{1}(\nu)/\nu$ of the P-wave in the low energy region.

 \section{Anomaly of the angular distribution of the \\
 neutron-Pb amplitude}
 
      Since the parameters of the long range component of the nuclear 
 force are already known by the analysis of the S-wave amplitude of 
 the proton-proton scattering, we can determine the potential of the 
 neutron-nucleus scattering, at least in the region of the long range 
 tail.  It must be emphasized that the singular behavior of the 
 amplitude at $t=0$ is determined solely by the asymptotic behavior 
 of the long range potential, and therefore the change of the potential 
 for finite $r$, for example inside of the nucleus, does 
 not have any effect on the singularity at $t=0$.
   All the information on the long range component of the nuclear 
 force is contained in the spectral function
 \begin{equation}
 A_{t}^{extra}(s,t)=\pi C' t^{\gamma} e^{-\beta t}
 \end{equation} 
 with 
 \begin{equation}
 \gamma=1.54 \quad , \quad C'=0.175 \quad and \quad \beta=0.0626
 \end{equation} 
 in the unit of the neutral pion mass.  The value of $\gamma$ is 
 close to that of the Van der Waals force of the London type(
 $\gamma=1.5$) and the 
 sign of $C'$ indicates that the force is attractive. 
 In this section, we shall compute the amplitude of the 
 neutron-Pb$^{208}$ scattering using the parameters $\gamma$ and $C'$,
 and predict the shape of the singular behavior at $t=0$.
 Such a prediction will be helpful in designing the neutron-Pb$^{208}$ 
 experiments of high precision and the data obtaind in the experiments 
 will in turn improve the values of the parameters $\gamma$ and $C'$.
 
     In computing the convolution of the nuclear potential and the 
 form factor $\rho (r)$ of the nucleus, for reason of simplicity, 
 we shall choose $\rho (r)$ as the box form of radius $r_{1}$, 
 namely $\rho (r)$ is zero and non-zero constant in $r>r_1$ and 
 in $r<r_1$ respectively.  This is permissible because we are 
 interested in the 
 singularity and it is not affected by the change of the potential 
 at finite $r$.  Among three parameters appeared in the spectral 
 function $A_{t}^{extra}(s,t)$ of Eq.(9), 
 $\beta$ does not relate to the 
 singularity, but it controls the depth of the n-Pb potential. 
 Contrary to $\gamma$ and $C'$, $\beta$ will be regarded as a free 
 parameter and whose value will be fixed by fitting to the scattering 
 length of the S-wave amplitude of the n-Pb scattering.    There is 
 another free parameter $r_{1}$, size of the nucleus.   We shall 
 consider two amplitudes, in which $r_{1}=4.4$ and $r_{1}=3.5$ 
 respectively.  The former is the standard size of the nucleus 
 of $A=208$, however $r_{1}=4.4$ gives the slope of $ 
 \sqrt{\nu} \cot \delta_{0} (\nu)$ around 10 \% higher.     If we 
 move $r_{1}$ to $3.5$ it gives right value $d(\sqrt{\nu} 
 \cot \delta_{0} (\nu))/d \nu=1.9$, which was obtained in the 
 transmission experiment of the n-Pb$^{208}$ scattering.
 In the following, we shall designate these two amplitudes 
 starting from the strong Van der Waals potential  [vdw44] 
 and [vdw35] respectively.  In order to compare the cases 
 of the long range forces with that of the short range force, 
 the third amplitude of n-Pb will be considered  starting from 
 the nuclear potential of the sigma meson exchange with $m_{\sigma}=4$. 
 In the calculation we shall choose 
  $r_{1}=4.4$ and the coupling constant $g_{\sigma}^2$ is 
 fixed by fitting to the scattering length of the S-wave.
 This amplitude will be designated  [sigma].   In figure 1, the 
 potentials $V(r)$ of the n-Pb scatterings are shown.

 \begin{figure}[htbp]
\includegraphics[width=.8\textwidth,height=7.0cm]{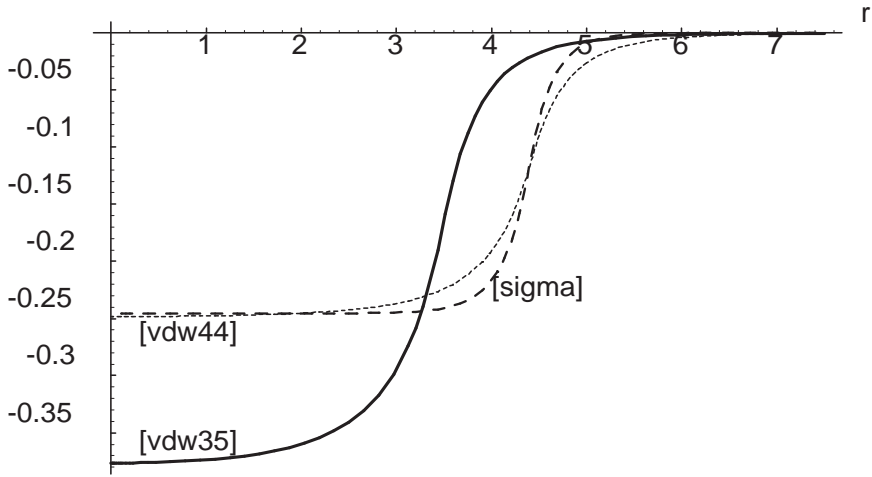}
\caption{{\footnotesize
The potentials of neutron-Pb interactions [vdw35], [vdw44] and 
[sigma] are plotted against $r$.  [vdw35] and [vdw44] are constructed 
from the nuclear potential of the strong Van der Waals type, whose 
radius of the nucleus $r_{1}$ is 3.5 and 4.4 respectively. Whereas
 [sigma] is constructed from the nuclear potential of $\sigma$-meson
  exchange, and $r_{1}=$4.4 . }}
\end{figure}
 
   In this paper, we shall consider the n-Pb scattering in 
 $T_{lab} \leq 1$ MeV..   Since $\nu$ and $T_{lab}$ is proportional 
 and $T_{lab} = 1$ MeV. corresponds to $\nu=0.102123$, the domain 
 of $-t$ necessary to make the patial wave projection is $0 \leq 
 -t <0.41$, whereas the nearest singularity on the $t$-plane 
 arising from the exchange of the sigma meson is $t=16$.  Therefore 
 we expect that the partial wave amplitudes $a_{\ell}(\nu)$ of [sigma] 
 decreases very rapidly with $\ell$.   On the other hand, for the case 
 of the Van der Waals interaction in which all the threshold 
 behavior is 
 proportional to $\nu^{\gamma}$ for $\ell \geq 2$, $a_{\ell}(\nu)$ 
 does not decrease so rapidly even if $\nu$ is small.   However we 
 can see that the difference of the partial wave amplitudes from those 
 of the Born term, namely $(a_{\ell}(\nu)-a_{\ell}^{(Born)}(\nu))$, 
 decreases very rapidly with $\ell$ for small $\nu$.   Therefore 
 the amplitudes of [vdw44] and [vdw35] are
 \begin{equation}
 F(\nu,t)=F^{Born}(\nu,t)+ \sum_{\ell =0}^{L} (2 \ell +1) 
 (a_{\ell}(\nu)-a_{\ell}^{(Born)}(\nu)) P_{\ell}(z) \quad ,
 \end{equation} 
 and we shall set $L=4$ because $|a_{5}(\nu)-a_{5}^{(Born)}(\nu)| 
 < 10^{-4}$. 
   Since the n-Pb potential $V(R)$ is the convolution of $\rho(r)$ 
 and $v(r)$, the Born term of the n-Pb scattering amplitude $F^{Born}
 (\nu,t)$ is the product of the Fourier transformations of these 
 functions.  
 
      The first one is 
 \begin{equation}     
 A \hat{\rho}(t) \equiv \int d^3 r e^{- i \vec{q} \cdot \vec{r}}
 \rho(r)=A \frac{3 (\sin x -x \cos x)}{x^3} |_{x=q r_1} \qquad ,
 \end{equation} 
  where $q$ is the momentum transfer and relates to $t$ by 
 $t=-q^2$.  Since $\hat{\rho}(t)$ is an even function of $q$, 
 it is a regular function of $t$, and it is normalized as 
 $\hat{\rho}(0)=1$.  The second one is the Born 
 term of the neutron-nucleon scattering:
 \begin{equation}
  A^{extra}(t) \equiv \frac{1}{\pi} \int dt' \frac{A_{t}^{extra}(s,t')}
  {t'-t} = \Gamma(\gamma+1) C' (-t)^\gamma e^{-\beta t}
   \Gamma (-\gamma, -\beta t,\infty) \quad ,
 \end{equation} 
 in which the incomplete gamma function is used and 
 whose definition is 
 \begin{equation}
 \Gamma(g,a,b)=\int_{a}^{b} dx x^{g-1} e^{-x}
 \end{equation} 
 $\Gamma (g,p,\infty)$ can be 
 divided into two terms: $(\Gamma(g)-\Gamma(g,0,p))$.   
 If we use the separation, $A^{extra}(t)$ is written as the sum of 
 the singular and the regular terms of $t$ :
  \begin{equation}
  A^{extra}(t) = - \frac{\pi C'}{\sin \pi \gamma}  
  (-t)^{\gamma} e^{-\beta t} -\Gamma (\gamma+1) \frac{C'}{\beta^\gamma} 
  (-\beta t)^{\gamma} e^{-\beta t} \Gamma(-\gamma,0,-\beta t) \quad .
  \end{equation} 
  From Legendre's formula
 \begin{equation}
 p^{-g} e^{p} \Gamma(g,0,p)= \sum_{n=0}^{\infty} \frac{p^n}{g (g+1)
 (g+2) \cdots (g+n)} \quad ,
 \end{equation} 
  we see that the second term of the r.h.s. of Eq.(15)
 is a regular function of $-\beta t$.
 By multiplying $A \hat{\rho}(t)$, we finally obtain the Born 
 term of the scattering amplitude $F^{Born}(\nu,t)$
 \begin{equation}
 F^{Born}(\nu,t)= F_{s}(\nu,t) (-t)^{\gamma} + F_{r}(\nu,t) \quad , 
 \end{equation} 
  where $ F_{s}(\nu,t)$ and $ F_{r}(\nu,t)$ are regular functions 
 of $t$, and are defind by
 \begin{equation}
  F_{s}(\nu,t)= - \frac{\pi}{m \sin \pi \gamma} A C' e^{-\beta t}
   \hat{\rho}(t) \quad ,
 \end{equation} 
 \begin{equation}
  F_{r}(\nu,t)= - \frac{A C'}{m \beta^{\gamma}} \Gamma(\gamma+1)
  \{ (-\beta t)^{\gamma} e^{-\beta t} \Gamma(-\gamma,0,-\beta t) \} 
  \hat{\rho}(t)
 \end{equation} 
 respectively.

     Contrary to the case of the short range force, the amplitudes 
 with the long range force cannot be fitted by a few terms of the 
 polynomials of $z$.   In fact curves [vdw35] and [sigma] 
 in the figures show  
 such property, where [vdw35] and [sigma] are the scattering 
 amplitudes $F(\nu,t)$ minus their S, P and D waves for the 
 Van der Waals force and for the $\sigma$-meson exchange force 
  respectively.
  Smallness of [sigma] means that the amplitude arising from 
 the $\sigma$-meson exchange is fitted well by the quadratic   
 function of $z$.   In order to confirm that the singularity in 
 [vdw35] is actually $F_{s}(\nu,t) (-t)^{\gamma}$, we must show the 
 smallness of $\{F(\nu,t)- F_{s}(\nu,t) (-t)^{\gamma}\}$ minus the 
 S, P and D  waves.   In the figure, the dotted curve 
 is such a regular part of [vdw35].   Figure 2 and 3 are the 
 graphs of $T_{lab}=$ 0.25 and 0.5MeV. respectively, in which 
 the full line is the [vdw35], the dashed line is the [sigma] 
 multiplied by 10 and the dotted line is the regular part of 
 [vdw35] multiplied by factor 100.

\begin{figure}[htbp]
\begin{minipage}{6.8cm}
\includegraphics[width=.99\textwidth,height=5.0cm]{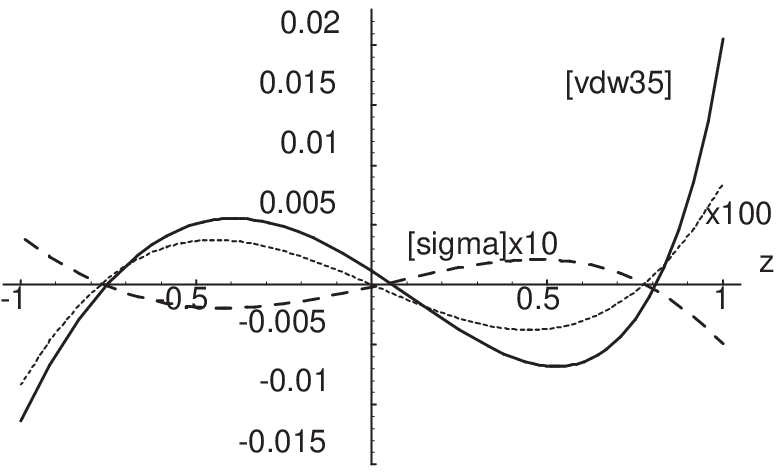}
\caption{{\footnotesize
$T_{lab}=$0.25MeV..  Amplitudes $F(\nu,t)$ minus the S, P 
and D waves are plotted against $z$. The dotted curve, which is 
the regular part of [vdw35], is multiplied by 100.}}
\end{minipage}
\hfill
\begin{minipage}{6.8cm}
\includegraphics[width=.99\textwidth,height=5.0cm]{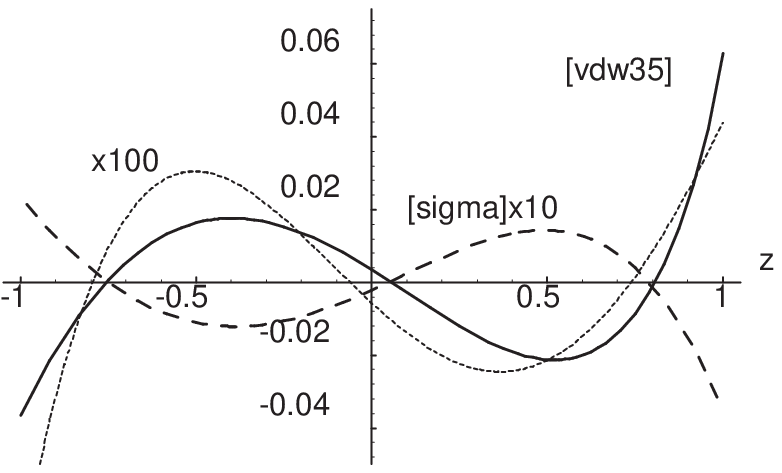}
\caption{{\footnotesize 
$T_{lab}=$0.50MeV..  Amplitudes $F(\nu,t)$ minus the S, P 
and D waves are plotted against $z$.  The dotted curve, which is 
the regular part of [vdw35], is multiplied by 100.}}
\end{minipage}
\end{figure}

  Figure 4 and 5 are the 
 graphs of $T_{lab}=$ 0.75 and 1.0MeV. respectively, in which 
 the contents are the same as fig.2 and fig.3 except the 
 multiplicative factor, namely in these figures only the dotted 
 curves are multiplied by factor 10.    As the incident energy 
 increases, even the amplitude arising from the short range 
 potential becomes more and more difficult to be fitted by the 
 quadratic function of $z$.   Therefore we must try the fit 
 by the cubic function for $T_{lab}=$ 0.75 and 1.0MeV..
   Figure 6 and 7 are the 
 graphs of $T_{lab}=$ 0.75 and 1.0MeV. respectively, in which 
 the contents are the same as fig.4 and fig.5 except the 
 subtracted functions are the cubic function, namely the S, 
 P, D and F waves, rather than the quadratic function of $z$.
   And  as in fig.4 and fig.5, only the regular part of 
   [vdw35] (dotted curve) is multiplied by factor 10. 
    When we draw curves, values of parameters of [vdw35] are
 \begin{equation}
 \gamma=1.54 \quad , \quad \beta=0.2517 \quad , \quad C'=0.175 \quad
 and \quad r_{1}=3.5 
 \end{equation} 
 , whereas the nuclear potential of the $\sigma$-meson exchange is 
 chosen as $v(r)=-0.59 e^{-4 r}/r$ and $r_{1}=4.4$ for [sigma].
     
 \begin{figure}[htbp]
\begin{minipage}{6.8cm}
\includegraphics[width=.99\textwidth,height=5.0cm]{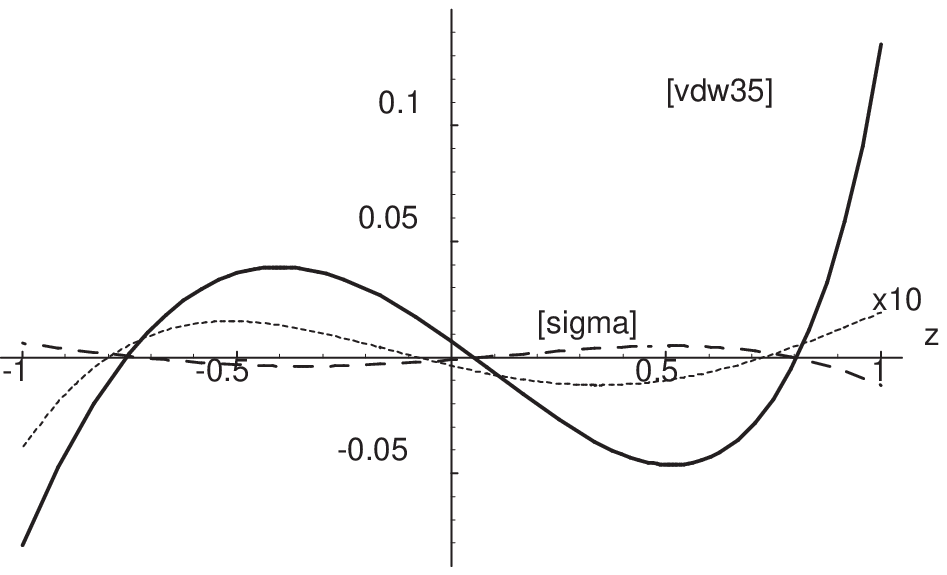}
\caption{{\footnotesize
$T_{lab}=$0.75MeV..  Amplitudes $F(\nu,t)$ minus the S, P 
and D waves are plotted against $z$. The dotted curve, which is 
the regular part of [vdw35], is multiplied by factor 10.}}
\end{minipage}
\hfill
\begin{minipage}{6.8cm}
\includegraphics[width=.99\textwidth,height=5.0cm]{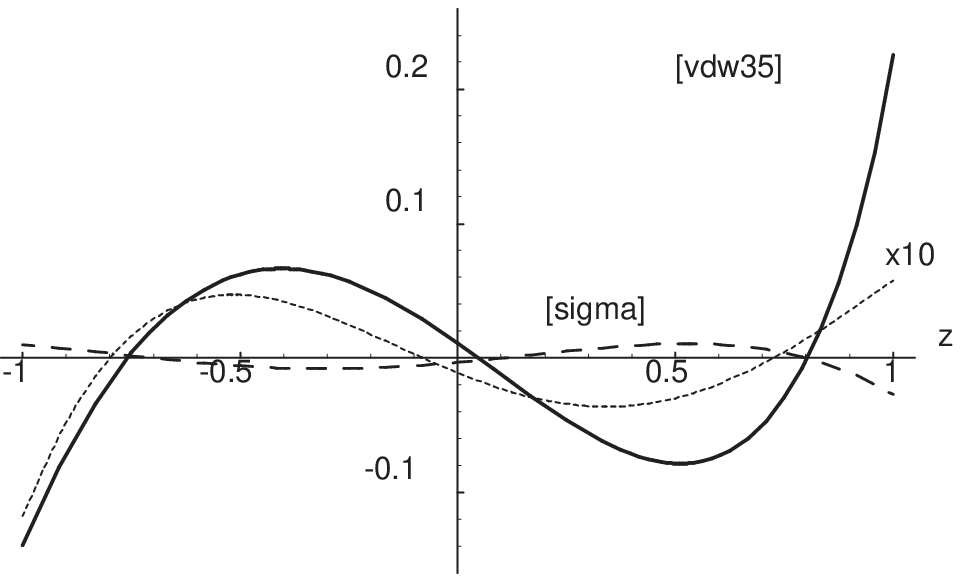}
\caption{{\footnotesize 
$T_{lab}=$1.0MeV..  Amplitudes $F(\nu,t)$ minus the S, P 
and D waves are plotted against $z$.  The dotted curve, which is 
the regular part of [vdw35], is multiplied by factor 10.}}
\end{minipage}
\end{figure}

 When the precise n-Pb experimants are carried out, 
 we shall obtain similar curves as [vdw35].    The content of 
 our prediction is that the singular behaviors of the experimental 
 curves will be removed by subtracting $F_{s}(\nu,t) (-t)^{\gamma}$ 
 with the parameters given in Eq.(20), which are obtained from the 
 data of the low energy p-p scattering.
  If the sufficiently accurate n-Pb data are available, we may expect 
 to obtain the improved values of the long range parameters 
 $\gamma$ and $C'$.
 
\begin{figure}[htbp]
\begin{minipage}{6.8cm}
\includegraphics[width=.99\textwidth,height=5.0cm]{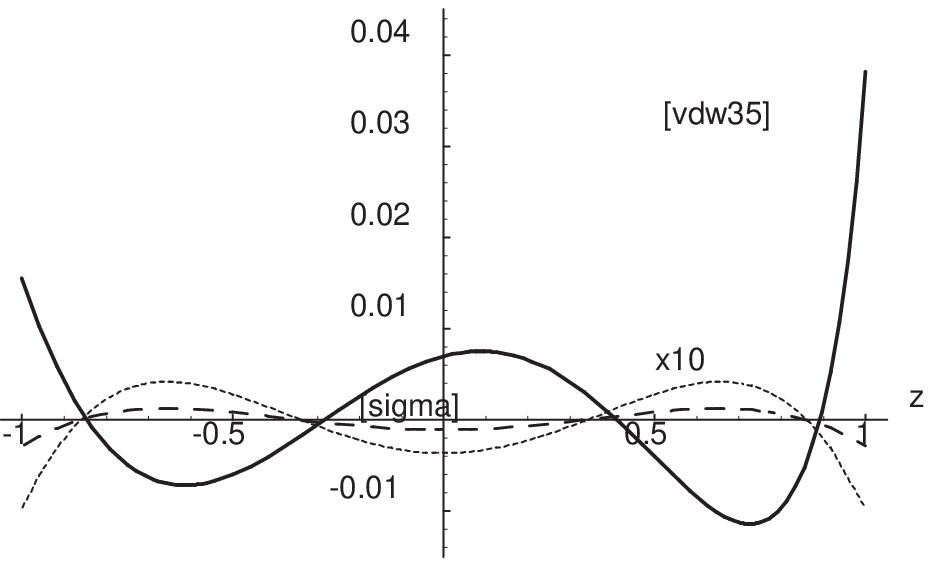}
\caption{{\footnotesize
$T_{lab}=$0.75MeV..  Amplitudes $F(\nu,t)$ minus the S, P,D 
and F waves are plotted against $z$. The dotted curve, which is 
the regular part of [vdw35], is multiplied by factor 10.}}
\end{minipage}
\hfill
\begin{minipage}{6.8cm}
\includegraphics[width=.99\textwidth,height=5.0cm]{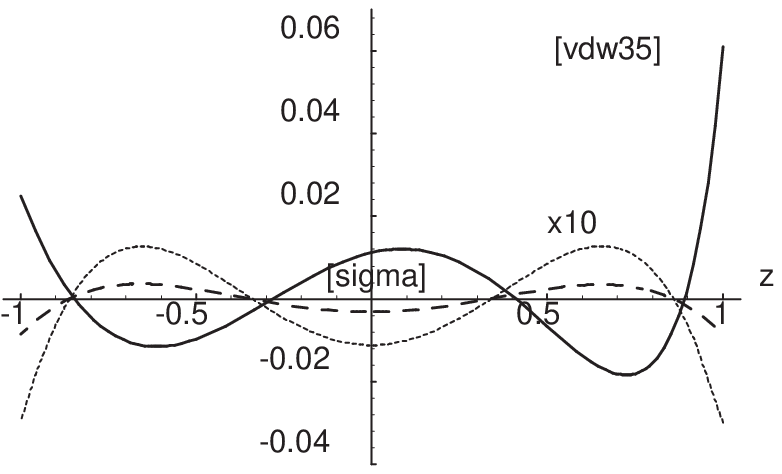}
\caption{{\footnotesize 
$T_{lab}=$1.0MeV..  Amplitudes $F(\nu,t)$ minus the S, P, D 
and F waves are plotted against $z$.  The dotted curve, which is 
the regular part of [vdw35], is multiplied by factor 10.}}
\end{minipage}
\end{figure}

 \section{ The cusp of $a_{1}(\nu)/\nu$ at $\nu=0$ }
 
       Since the singular term $(-t)^{\gamma}$ is equal to  $2^{\gamma}
   \nu^{\gamma} (1-z)^{\gamma}$, the singular behaviors appear in two 
  places.   One is the singular behavior $(1-z)^{\gamma}$ 
 of the angular distribution at $z=1$ for fixed $\nu$, and the 
 other is the singular term of $\nu^{\gamma}$ in the partial wave 
 amplitude.   In the previous section we studied the former, and the 
 latter will be studied in this section.  When $\gamma=1.54$, $\nu^
 {\gamma}$ is the leading threshold behavior of the D and the higher 
partial waves rather than the standard threshold behavior $\nu^{\ell}$. 
For the P-wave $a_{1}(\nu)$, the threshold behavior is proportional 
to $\nu$, and therefore we can use the once subtracted form 
$a_{1}(\nu)/\nu$.     In $a_{1}(\nu)/\nu$ the singular term becomes 
$\nu^{\gamma-1}$, and which is a cusp at $\nu=0$ because its slope is 
infinity.   When the long range tail of the nuclear potential is 
attractive, from Eqs.(7) and (8) the coefficient of 
$\nu^{\gamma-1}$ is negative and therefore the 
cusp must point upward.  On the other hand, the curve of [sigma] 
must be regular at $\nu=0$.
  In figure 8, such curves of [vdw35], 
[vdw44] and [sigma] are shown.  In order to make the display compact, 
since we do not have interest in the back ground regular functions, 
 we subtract linear functions, which are chosen in such 
ways that curves come to $\nu$-axis as close 
as possible in $0.02 <\nu <0.03 $.

\begin{figure}[htbp]
\includegraphics[width=.8\textwidth,height=8.0cm]{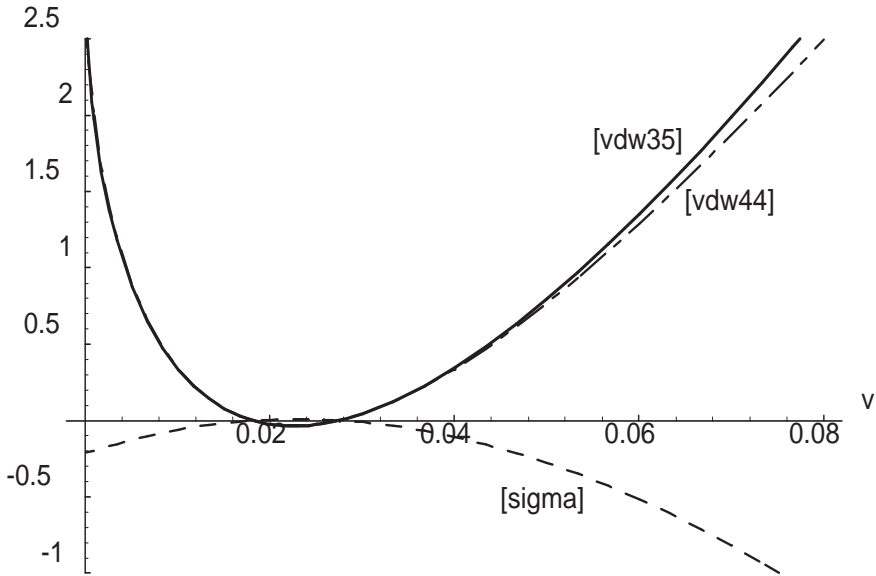}
\caption{{\footnotesize
The Kantor amplitudes of the P-wave $K_{1}(\nu)/\nu$ are plotted 
against $\nu$.  Curves [vdw35] and [vdw44] arising from the strong 
Van der Waals force have cusps at $\nu=0$, which point upward.  On 
the other hand, the curve [sigma] arising from short range force 
is regular at $\nu=0$.}}

\end{figure}

 If we speak precisely, the graphs in fig.8 are not the real parts of 
$a_{1}(\nu)/\nu$ but the once subtracted Kantor amplitude defined 
by\cite{kantor} 
\begin{equation}
\frac{K_{1}(\nu)}{\nu} = {\rm Re}\frac{a_{1}(\nu)}{\nu}- \frac{P}{\pi} 
\int_{0}^{\infty} d \nu' \frac{{\rm Im} a_{1}(\nu')}{\nu'(\nu'-\nu)}
\quad ,
\end{equation} 
 where $P$ stands for Cauchy's pricipal value integration.   It is 
important that the unitarity cut in $0 \leq \nu < \infty$ of 
$a_{1}(\nu)/\nu$ is removed in $K_{1}(\nu)/\nu$.  Therefore 
$K_{1}(\nu)/\nu$ has a wide domain of analyticity, and which is 
necessary to observe the delicate difference of behaviors arising 
from the spectrum on the left hand cut $-\infty < \nu \leq m_{1}^{2}
/4$ of the short range force, where $m_{1}$ is the lightest mass 
exchanged, and from the spectrum of the long range force in $-\infty 
< \nu \leq 0$.   It is fortunate that, for the P and higher partial 
waves, 
the integrations of Eq.(21) are small and in their estimation, very 
precise phase shift data are not required.    This is because the 
 threshold behavior of ${\rm Im} a_{1}(\nu')/\nu'$ is $\nu^{3/2}$ 
even for the P-wave.   On the other hand for the S-wave, the 
threshold behavior 
of ${\rm Im} a_{0}(\nu')/\nu'$ is $1/\sqrt{\nu}$ and the estimation 
of the principal value integration requires very precise data of 
the phase shift $\delta_{0}(\nu)$.  Another difficulty to use the 
S-wave amplitude is that we need to know extremely precise value 
of the scaatering length $-a_{0}(0)$, since in 
cmputing the once subtracted Kantor amplitude, we must know 
$(a_{0}(\nu)-a_{0}(0))/\nu$.  Therefore the S-wave is not the 
suitable place to observe the singularity $\nu^{\gamma-1}$, unless 
we can overcome these difficulties.    If we consider that we 
do not have sufficient data of the D and higher partial 
waves in the low energy region, the P wave is the most suitable 
place to observe the singularity $\nu^{\gamma}$ of the partial 
waves. 
 
 \section{Remarks and Comments}
    
    Since the Van der Waals force is universal\cite{universal}
    , its existence 
    in the nucleon-nucleon scattering implies the Van der Waals 
 interaction also in other processes such as in the pion-pion\cite
{pipi}  and in 
 the pion-nucleon scatterings.   The possibility to find such an 
 extra force depends on the precision of the data and the possibility 
 to prepare the wide domain of analyticity.   The low energy 
 data of the p-p scattering is prominent in their accuracy, whereas 
 the pion-pion process is prominent in its possibility to prepare 
 the wide analytic domain, because the two-pion exchange spectrum can be 
 constructed from the pi-pi amplitudes themselvs.   In the Appendix, 
 we see a cusp at $\nu=0$ in the once subtracted Kantor amplitude 
 of the S-wave of the p-p scattering, in which the one-pion exchange 
 spectrum is removed.   Similarly the characteristic cusp at $\nu=0$ 
 is observed in $a_{1}(\nu)/\nu$ of the pion-pion scattering, in 
 which the unitarity cut and the cut of the two-pion exchange are 
 removed.\cite{pipi}
 
     The third place easy to observe the strong Van der Waals force is 
 the low energy neutron-nucleus amplitude, because the strength $C$ 
 of the strong Van der Waals potential of the nuclear force 
 is magnified by factor $A$, 
 the mass number.  Although the effect of the Van der Waals force 
 decreases as $\nu^{\gamma}$ for small $\nu$, in the neuton-nucleus 
 scattering the large value of the strength $A C$ allows us to 
 observe it even at the smaller energy.  It is advantageous to use 
 the lower energy data in the determinations of the paremeters 
 $\gamma$ and $C'$ of the threshold behavior appeared in the extra 
 spectrum $A_{t}^{extra}(s,t)=\pi C' t^{\gamma} e^{-\beta t}$, 
  because they are determined without being disturbed by the 
  parameter $\beta$, which specifies the deviation from the 
  threshold behavior. 
    
       Therefore our proposal in this paper is to measure the 
  angular distribution of the cross section of the neutron-nucleus 
  scattering such as n-Pb$^{208}$ precisely for fixed energy 
  around $T_{lab}=$1MeV., and to determine the parameters of the 
  asymptotic tail of the long range component of the nuclear 
  potential.     Although the values of such parameters are known 
  from the low energy proton-proton data, the independent 
  determination  will serve to confirm the actual existence of 
  the strong Van der  Waals interaction in the nuclear force.      
 
 \vspace{5mm}
 
\begin{flushleft}
{\Large\bf Appendix}
\end{flushleft}
 
         When we extract the information of the long range force 
 from the partial wave amplitude $h_{\ell}(\nu)$, first of all we 
 must remove the unitarity cut, and then the known near-by 
 singularities.  In this way we can prepare a function with 
 the wide domain of analyticity in the neighborhood of $\nu=0$, 
 and which is helpful to observe the extra singularity at $\nu=0$
  arising from the long range interaction, if it exists.   Since 
 the phase shift data of the S-wave of the low energy proton-proton 
 scattering is most accurate in the hadron physics, we shall 
 consider the once subtracted S-wave amplitude $(h_{0}(\nu)-      
 h_{0}(0))/\nu$.   If we make the Kantor amplitude by
 \begin{equation}
 K_{0}^{once}(\nu)={\rm Re}(h_{0}(\nu)- h_{0}(0))/\nu
 -\frac{1}{\pi} \int_{0}^{\infty} \frac{{\rm Im} h_{0}(\nu')}
 {\nu' (\nu'-\nu)} d \nu'
 \end{equation} 
 , the unitarity cut is removed.  Since the one-pion exchange 
 (OPE) contribution is 
 \begin{equation}
  h_{0}^{1 \pi}(\nu)=\frac{1}{4} \frac{g^2}{4 \pi} \frac{1}{4 \nu}
  \log (1+4 \nu)
  \end{equation} 
 , which has a left hand cut starting from $\nu=-1/4$, the OPE cut 
 in the Kantor amplitude is removed in
  \begin{equation}
\tilde{K}_{0}^{once}(\nu)=K_{0}^{once}(\nu)
 -\frac{(h_{0}^{1 \pi}(\nu)-h_{0}^{1 \pi}(0))}{\nu} \quad .
  \end{equation} 
  
 When we have precise phase shift data, we can evaluate 
 $-\tilde{K}_{0}^{once}(\nu)$, and it is shown in figure 9 and 10. 
Although the spectrum of the two-pion exchange starts at $\nu=-1$, 
in the threshold region the spectrum is very small, because it 
arises from the partial wave projection of the continuous spectrum 
 $A_{t}^{2 \pi}(s,t)$.  If we replace $A_{t}^{2 \pi}(s,t)$ by the 
delta function of the $\sigma$-meson exchange, the left hand cut of 
$-\tilde{K}_{0}^{once}(\nu)$ starts at $\nu=-4$.  Therefore if the 
nuclear force is short range due to the exchanges of a set of pions,
 $-\tilde{K}_{0}^{once}(\nu)$ must be almost constant with very 
 small slope and extremely small curvature in the neighborhood of 
 $\nu=0$. 
 
 \begin{figure}[htbp]
\begin{minipage}{8.cm}
\includegraphics[width=.99\textwidth,height=5.5cm]{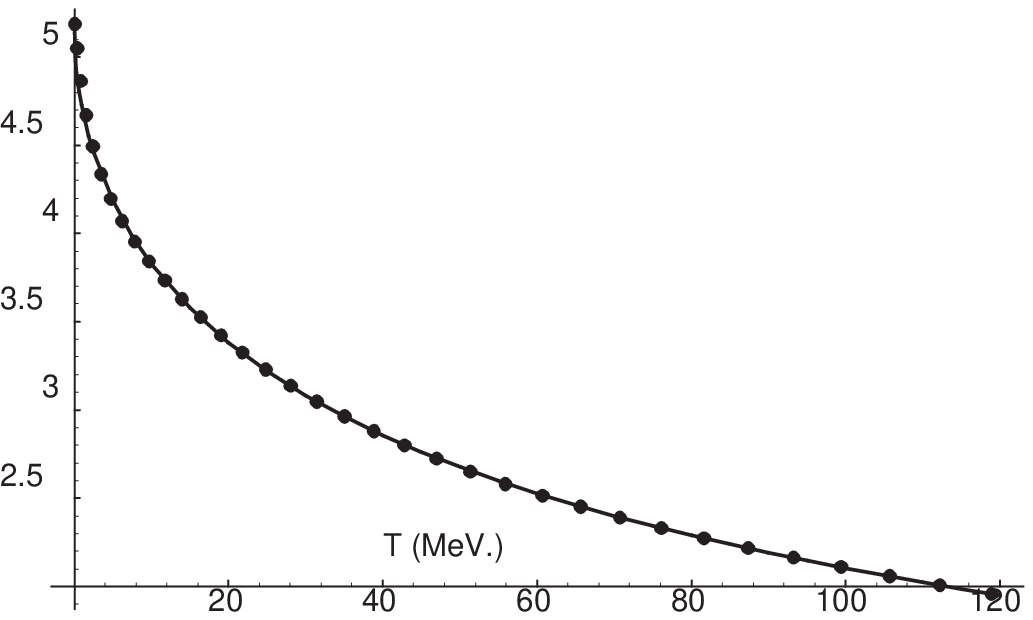}
\caption{{\footnotesize
 $-\tilde{K}_{0}^{once}(\nu)$ is plotted against $T_{lab}$ (MeV.). 
 The closed circles are evaluated from the S-wave phase shifts of 
 the p-p scattering.  The error bars are smaller than the size of 
 the circles in $T_{lab}>$4MeV..  The curve is the results of the 
 fit, in which the spectrum with three free parameters of Eq.(25)
 is searched. 
}}
\end{minipage}
\hfill
\begin{minipage}{5.8cm}
\includegraphics[width=.99\textwidth,height=5.8cm]{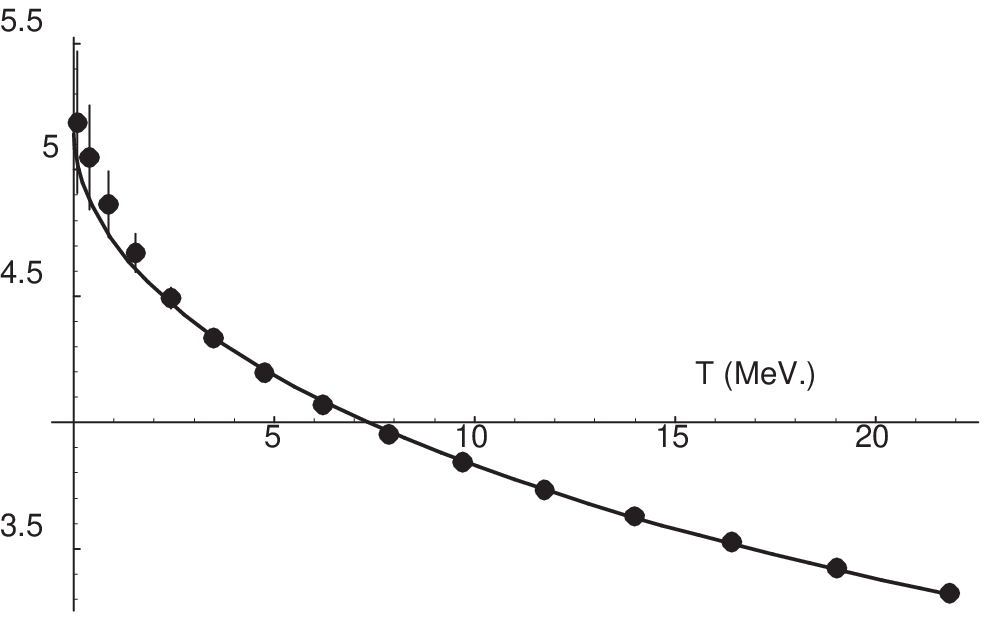}
\caption{{\footnotesize 
 The same graph in the lower energy region as the figure 9 is shown.
 The error bars are drawn.   The cusp at $\nu=0$, which points 
 upward, indicates the
 strong long range force with the attractive sign. 
}}
\end{minipage}
\end{figure}

  However figures 9 and 10 indicate this is not the case, 
 but there is a cusp at $\nu=0$ (closed circles).   This cusp is 
fitted by an extra spectrum of the long range interaction of the form 
\begin{equation}
   A_{t}^{extra}(4 m^2,t)= \pi C' t^{\gamma} e^{-\beta t} \quad .
\end{equation} 
More explicitely, it is fitted by
\begin{equation}
\frac{h_{0}^{extra}(\nu)-h_{0}^{extra}(0)}{\nu}=2 C' \int_{0}^{\infty}
dt \, t^{\gamma} e^{-\beta t} \{ \frac{1}{2 \nu} Q_{0}
(1+\frac{t}{2 \nu})
-\frac{1}{t} \} \frac{1}{\nu}
\end{equation} 
, in which the extra factor 2 in front of the integration appears 
because the contributions from the spectrum $A_{u}$ as well as from 
$A_{t}$ must be included.   The integration of Eq.(26) can be written 
in terms of the generalized hypergeometric function $_{p} F_{q} (x)$ 
and of the confluent hypergeometric function $F(a,b,x)$.  
\begin{equation}
\frac{h_{0}^{extra}(\nu)-h_{0}^{extra}(0)}{\nu}=
2 (\nu^{\gamma -1} \xi_{s}(\nu)+\xi_{r}(\nu))
\end{equation} 
, where $\xi_{s}(\nu)$ and $\xi_{r}(\nu)$ are regular functions of 
$\nu$ and are difined by
\begin{equation}
\xi_{s}(\nu)=-C' \frac{4^{\gamma}}{\gamma +1} \frac{\pi}{\sin \pi 
\gamma} F(1+\gamma,2+\gamma,4 \beta \nu)
\end{equation} 
and
\begin{eqnarray}
\xi_{r}(\nu)
 &=& C' \frac{\Gamma(\gamma)}{\beta^{\gamma} \nu} 
(_{\{1,1 \}} F_{\{2, 1-\gamma \}}(4 \beta \nu)-1) \nonumber \\
 &=&
C' \frac{\Gamma(\gamma)}{\beta^{\gamma} \nu}  ( \sum
_{n=1}^{\infty} \frac{(4 \beta \nu)^{n}}{(1-\gamma) \cdots (n-\gamma)}
\frac{1}{n+1} )
\end{eqnarray} 
respectively.
  Results of the chi-square fits are 
\begin{equation}
\gamma=1.543 \quad , \qquad \beta=0.06264 \quad , \qquad 
C'=0.1762 \qquad and \qquad \chi=0.441
\end{equation} 
in the unit of the neutral pion mass.  Among three parameters,
 $\gamma$ and $C'$ are the threshold parameters of the long range 
 force, whereas $ \beta$ is necessary to make the integration 
 convergent, and $\sqrt{\beta}$ must be the order of magnitude of 
 the nucleon radius.   In constructing the neutron-Pb potential, 
 although we use the values of $\gamma$ and $C'$ given in Eq.(30),
  $\beta$ is left as a free parameter and later fixed by fitting 
  to the scattering length of the n-Pb amplitude.    Up to this 
  point we have not considerd the ordinary Coulomb interaction
 and the vacuum polarization.    Therefore the procedure explained
 above is applicable only to the neutron-neutron data. \ \  When we 
 include the electromagnetic interaction, some modifications are 
 necessary in the constructions of the singularity free functions 
 $K_{0}(\nu)$ from the phase shift functions.   
 An explicit construction of the Kantor amplitude 
 from the phase shift $\delta_{0}^{E}(\nu)$ are explained in other 
 paper.\cite{ppvdw}
 
     Finally it will be helpful to write the normalization of the 
 amplitudes explicitly:    
 \begin{equation}
 A(s,t)=\sum_{\ell=0}^{\infty} (2 \ell +1) h_{\ell}(\nu) P_{\ell}
 (z) \quad with \quad h_{\ell}(\nu) = \frac{\sqrt{m^2+\nu}}{
 \sqrt{\nu}} e^{i \delta_{\ell}(\nu)} \sin \delta_{\ell}(\nu)
 \end{equation}
 and
 \begin{equation}
 f(\nu,t)=\sum_{\ell=0}^{\infty} (2 \ell +1) a_{\ell}(\nu) P_{\ell}
 (z) \quad with \quad a_{\ell}(\nu) = \frac{1}{
 \sqrt{\nu}} e^{i \delta_{\ell}(\nu)} \sin \delta_{\ell}(\nu) \quad .
 \end{equation}
 Therefore $f(\nu,t)=A(s,t)/m$ in the approximation $\nu \ll m^2$.

 \end{document}